\documentstyle[12pt]{article}
\begin{document}
\baselineskip=0.7cm
\newcommand{\EQ}{\begin{equation}}
\newcommand{\EN}{\end{equation}}
\newcommand{\EQA}{\begin{eqnarray}}
\newcommand{\EQN}{\end{eqnarray}}
\newcommand{\e}{{\rm e}}
\newcommand{\Sp}{{\rm Sp}}
\renewcommand{\theequation}{\arabic{section}.\arabic{equation}}
\newcommand{\Tr}{{\rm Tr}}
\renewcommand{\thesection}{\arabic{section}.}
\renewcommand{\thesubsection}{\arabic{section}.\arabic{subsection}}
\makeatletter
\def\section{\@startsection{section}{1}{\z@}{-3.5ex plus -1ex minus
 -.2ex}{2.3ex plus .2ex}{\large}}
\def\subsection{\@startsection{subsection}{2}{\z@}{-3.25ex plus -1ex minus
 -.2ex}{1.5ex plus .2ex}{\normalsize\it}}
\makeatother
\def\thefootnote{\fnsymbol{footnote}}
\begin{flushright}
NSF-ITP-93-67\\
BROWN-HEP-904\\
UT-KOMABA/93-10 \\
hep-th/9305109\\
May 1993
\end{flushright}

\begin{center}
\Large
A Deformed Matrix Model and the Black Hole Background in Two-Dimensional
String Theory
\\

\vspace{1cm}

\normalsize
{\sc Antal Jevicki}
\footnote{
e-mail address:\ \ antal@het.brown.edu}
\\
\vspace{0.3cm}
{\it Institute for Theoretical Physics, University of California,
Santa Barbara\\}
and\\
{\it Department of Physics, Brown University, Providence}
\footnote{
Permanent address}
\\

\vspace{0.5cm}

{\sc Tamiaki Yoneya}
\footnote{
e-mail address:\ \ yoneya@tkyvax.phys.s.u-tokyo.ac.jp {\it or}
 yoneya@sbitp.ucsb.edu \ (until June 30, '93)}
\\
\vspace{0.3cm}
{\it Institute for Theoretical Physics, University of California,
Santa Barbara\\}
and\\
{\it Institute of Physics, University of Tokyo, Komaba, Tokyo}
\footnote{
Permanent address}

\vspace{1.3cm}
Abstract\\
\end{center}
We invesigate how the exact 2D black-hole solution for the
critical string theory should be described, at least perturbatively
with respect to the inverse mass of the black hole, within
the framework of matrix model. In particular, we propose
a working hypothesis on the basis of which we can
present plausible candidates for the necessary
non-local field redefinition of the tachyon field and
the deformation of the usual $c=1$ matrix model
with $\mu =0$. We exhibit some marked difference
in the properties of tachyon scattering of the deformed model
from those of the usual $c=1$  model corresponding to tachyon
condensation. These results lead to
a concrete proposal for the S-matrix for the tachyon
black-hole scattering.

\newpage
\section{Introduction}
We now have some sound evidences for the fact that the standard
hermitian matrix model in one-demensional target space can be
regarded in the double scaling limit as a solution of
critical string theory \cite{Klebanov}. Namely, it corresponds to a
special solution of the $\beta$ function condition
with the condensation of tachyon and dilaton in a flat
two-dimensional target space time.
One of the great merits of matrix-model formulation of the
critical string theory will be that it is in principle
defined non-perturbatively with respect to string-coupling constant.
Furthermore, it is really remarkable that $W_{\infty}$ symmetry
structure associated with the ground ring in the CFT
approarch emerges in a simple and natural way associated
with an algebraic structure of the
special inverted oscillator hamiltonian of the matrix model
\cite{Jevicki}.

There has been, however, a basic question remaining to be
understood in the matrix model approach to critical
string theory. In the CFT approach, one has  an exact classical
solution describing a black-hole solution\cite
{Witten1}\cite{Mandal} to the critical string theory
which is different from a solution with tachyon condensation mentioned
above. In the matrix model approach, however, we have not established
yet whether and how, if any, the black hole background can be
included in the framework.
Witten\cite{Witten1}
 has proposed that the $c=1$ matrix model should be interpreted as an
extreme limit of the black hole in the sense that the black hole
horizon recedes to
infinity. This interpretation is indeed natural at least in
the limit of vanishing tachyon condensation, or in terms of
the 2D gravity picture, zero cosmological constant. But it raises
a question about what is then the non-exteme black hole solution
with finite mass.
If matrix models in general were to be trully  representations of the
solutions of critical string theory
as we hope, there should exit a matrix model
corresponding to the non-extremal black hole, since it
must reproduce the genus expansion in the weak coupling regime and
the black hole backgound is a perfectly good exact solution at
zero genus.

The purpose of this paper is to present some of our recent
efforts
\footnote{
For a very preliminary report, see \cite{Yoneya}
which contains some of the initial materials for the present work.
}
towards the identification of a matrix model describing the black hole
background.
Although our results are not yet completly conclusive, we believe that
our various observations already contain enough materials which
help to clarify the issues and to establish the foundations for
 further investigations.
\footnote{
There are also related works
\cite{Martinec}\cite{Das}\cite{Dhar}\cite{Russo}\cite{Yang}
 from which we however differ at
various points.}

To make the paper reasonablly self-contained and to
explain our view point, we begin
the next section with a brief review on the
spacetime interpretation of the ususal $c=1$ matrix model and the
properties of the black hole solution. We emphasize a trivial but
crucial relationship between the mass of a black hole and
the string  coupling constant. Based on this review, we propose
 a working hypothesis in section 3.
The hypothesis consists of two key ingredients, namely  a non-local
field redefinition of the tachyon field
and a deformation from the usual $c=1$ matrix model
at  $\mu = 0$. In section 4, we then present
our candidate for the non-local field redefinition. In section 5, we discuss
a plausible candidate for the deformed matrix model.
The model has some peculiar properties which are markedly different from the
usual $c=1$ model and deserves for further study independently
of the problem of black hole.
We derive  on-shell scattering amplitudes for the deformed
model in the tree approximation. In the final section, we discuss  a candidate
for $S$-matrix elements for black hole and tachyon scattering in
weak coupling perturbation theory and conclude by indicating the
limitation of the present work and future perspective.

\section{Space-Time Interpretation of the 1D
Matrix Model and the Black-Hole Solution}
\subsection{$c=1$ matrix model and its spacetime interpretation}

Why and how a one-dimensional
matrix model embodies string theory with  2-dimensional
target space is very mysterious and still remaines to be clarified.
But, the  reason for  a two dimensional target space to appear is not
difficult to see. Let us start from the standard double scaled
Lagrangian for an hermitian
matrix $M(t)$ in one dimension,
\EQ
L(M) = \Tr ( { 1 \over 2} \{ \dot M^2 - M^2 \}).
\label{matlag}
\EN
A convenient description of this system is in terms of a collective
field which is essentially a density function,
$\rho (x,t) = \Tr \delta (x - M(t))$
 of the
eigenvalues of $M$. The hamiltonian for the collective field
\cite{JS}\cite{DJ} is given by
\EQA
H &=& -\int dx [ {1 \over 2} \partial_x P_{\rho} \rho \partial P_{\rho}
+  {1 \over 6}\pi^2 \rho^3 + (-{1 \over 2}x^2 + \mu)\rho]\\
&=& {1 \over 2\pi} \int dx \int_{\alpha_-(x)}^{\alpha_+(x)} dp h(p,x),
\label{collham}
\EQN
where $h(p,x)=(p^2-x^2)/2+\mu$ is the one-body hamiltonian corresponding to
the free fermion picture for the matrix model, and $P_{\rho}$ is the
conjugate momentum of the collective field.
The parameter $-\mu$ is  the fermi energy. The new fields $\alpha_{\pm}$
are combinations $\partial_x P_{\rho} \mp \pi \rho $, satisfying
Poisson bracket relations
$\{\alpha_{\pm}(x), \alpha_{\pm}(x')\}= \mp 2\pi\delta'(x-x')$
. Polchinski\cite{Pol2} showed that in the semi-classical approximation,
the variables $\alpha_{\pm}$ can be interpreted as the branches of profile
function for the fermi surface in the classical phase space $(p,x)$.

The hamiltonian (\ref{collham}) can be regarded as describing a field theory
in a two-dimensional  target spacetime $(t, x)$. The equation of motion is
essentially a Liouville equation which in terms of $\alpha_{\pm}$ reads
\EQ
\dot \alpha_{\pm}= -\partial_x h(\alpha_{\pm},x).
\label{alphaeq}
\EN
The classical ground state corresponds to a static solution
\EQ
\alpha_{\pm}=\pm \sqrt{x^2-2\mu}\,\theta(x^2-2 \mu)\equiv \pm \alpha_0.
\label{clagr}
\EN
If we further redefine a new shifted field $\zeta$ and its conjugate
$\Pi_{\zeta}$
by
\EQ
\alpha_{\pm}= \pm \alpha_0 + \sqrt{\pi}({dx \over d\sigma})^{-1}
(\Pi_{\zeta}\mp \partial_{\sigma} \zeta)
\label{shift}
\EN
and a parametrized coordinate $\sigma$ by $x= \sqrt{2\mu} \cosh \sigma$
satisfying $dx/d\sigma = \alpha_0$, the hamiltonian (\ref{collham})
takes a more familiar looking form,
\EQ
H=\int_0^{\infty} d\sigma [{1 \over 2}(\Pi_{\zeta}^2 +
 (\partial_{\sigma}\zeta)^2)
+{\sqrt{\pi} \over 12} ({dx \over d\sigma})^{-2}\{ (\Pi_{\zeta}-\partial_
{\sigma}\zeta)^3-(\Pi_{\zeta}+\partial_{\sigma}\zeta)^3\}]
\label{collham2}
\EN
where we droped a c-number contribution. The boundary conditon for $\zeta$
at $\sigma = 0$ is known to be the Dirichret condition $\zeta (t, 0)=0$.
Thus, the system is interpreted as a local field theory for a massless scalar
field $\zeta$ with cubic interaction terms, whose strength is spatially
varying, being proportional to
\EQ
({dx \over d\sigma})^{-2}= {1 \over 2\mu\sinh^2 \sigma}.
\label{couling}
\EN

Here we emphasize that the emergence of linearized massless field is
actually not a consequence of a specific form of the potential $V(M)$
of the matrix model. The massless nature of $\zeta$ is
universal for aribitrary
potential provided ground state solution is static. Only the coupling
function depends on the form of potential.

The spatial dependence of coupling strength and  the massless
nature  of the scalar field suggest to identify  $\zeta$ to the
massless tachyon field in the so-called linear dilaton vacuum
which is an exact classical  solution of 2D critical string theory.
The linear dilaton vacuum corresponds to the classical background fields
\EQ
G_{\mu\nu}= \eta_{\mu\nu} \label{metric},
\EN
\EQ
\Phi = 2\sqrt{2} \phi \label{dil},
\EN
\EQ
T(x)=0 \label{tach},
\EN
for the world sheet action
\footnote{
We use the unit $\alpha' = 2$ which seems to be more-or-less standard
in the literature of conformal field theory.}
\EQ
S={1 \over 8\pi}\int d^2\xi (G_{\mu\nu}
\partial X^{\mu}\partial X^{\nu} - \sqrt{2} R^{(2)}\Phi (X)
+ 2 T(X))
\label{worldaction}
\EN
where we used a two-dimensional notation for the
 coordinates $X^{\mu}=(t, \phi)$ of the target Minkowski spacetime.
The physical spectrum around this background consists of
massless tachyon which is the only propagating degree and additional
infinite  number of ``discrete states''\cite{Polyakov}
\cite{GKN} with discrete
imaginary momenta and energies as given by
$ip_{\phi}= -\sqrt{2}(1-j),\, p_t=-i\sqrt{2}\,m$ with
$j= 1/2, 1, 3/2, 2, \ldots$ and $m= -j, -j+1,\ldots, j$.
It is well known that the discrete states generate an infinite
symmery algebra $W_{\infty}$.

The linearized equation for the tachyon fluctuation is
the zero mode part of the Virasoro condition
\EQ
L_0 T\, (=\bar L_0 T)\,=T, \label{viracon}
\EN
\EQ
L_0= {1 \over 2}(\partial_t^2 - \partial_{\phi}^2)-\sqrt{2}\phi.
\label{viraoplinear}
\EN
The redefined tachyon field $\tilde T \equiv \e^{\sqrt{2}\phi}T$
satisfies the massless Klein-Gordon equation. On the other hand, the
coupling strength of strings is
\EQ
g_{{\rm st}} = \e ^{-\Phi /2}=\e ^{-\sqrt{2}\phi}.
\label{stringcoupl}
\EN
Thus, at least asymptotically, we can make following
identification of the coordinates and fields,
\EQA
\sigma &\leftrightarrow& \phi/\sqrt{2},\label{dic11}\\
t_0    &\leftrightarrow& t/\sqrt{2},\label{dic12}\\
\zeta(t_0, \sigma) &\leftrightarrow& \tilde T(t, \phi)\label{dic13}.
\EQN
Here and what follows we always denote $t_0$ as the time coordinate of
the matrix model. Under this identification,  the appearence of
a scaling parameter $\mu$ can be naturally explained
in the  framework of critical string theory by assuming the presence
of a tachyon condensation of the form
\EQ
T(x) \sim \mu \e^{-\sqrt{2}\phi},
\label{tachycond}
\EN
which corresponds to a static solution of the linearized tachyon
equation.
(More precisely, one has to take account of the degeneracy of the static
solutions as discussed in ref. \cite{Pol1}.)
In particular, that the string coupling constant has a constant factor
$g_{{\rm st}} \propto \mu^{-1}$ is a
direct consequence of the scaling relation with respect to a
constant  shift of the coordinate $\phi$.

Now under the dictionary (\ref{dic11}) and (\ref{dic12}) for the
coordinates, the correspondence of the momentum and energy is given by
\EQA
i p_{\phi}& =& -\sqrt{2} + {1 \over \sqrt{2}} ip_{\sigma}, \label{dic21}\\
i p_t &=& i{1 \over \sqrt{2}}p_{t_0} \label{dic22}.
\EQN
The values of momentum and energy of the discrete states are then
$\, ip_{\sigma}=-2j, \,ip_{t_0}= -2m$, in terms of matrix model. There
indeed exit special operators with precisely these values of
momentum and energy,
\EQ
A_{jm}\equiv ({p + x \over 2})^{j+m}({p-x \over 2})^{j-m}.
\label{eigenop}
\EN
These are eigenoperators for the single-body hamiltonian
$h(p,x)$ with purely imaginary eigenvalue $-2im$ and generate a
$W_{\infty}$ algebra\cite{AJ1,MPY,MS,Witt2,KP,DDMW}
. This implies that the Schr\"{o}dinger equations
and hence the collective field theory is invariant under
special time dependent canonical transformation generated by
$\e^{-2mt_0}A_{jm}$.

The emergence of the $W_{\infty}$ structure associated with
the discrete states strengthens the above correspondence of
matrix model and the critical string theory, since the
discrete states are really the remnants of excitation modes of
critical strings in higher dimensions.
There is, however, a small puzzle here. The linearized equation for the
scalar quanta $\zeta$  is valid even for finite
$\sigma$, whereas the free massless tachyon equation (\ref{viracon})
with (\ref{viraoplinear}) of string theory is only valid
in the asymptotic region where the tachyon condensation (\ref{tachycond})
is neglected. When the condensation of tachyon is taken into account,
the Virasoro condition is expected to be modified into
, at least for sufficiently large $\phi$,
\footnote{
The result of section 7 of ref. \cite{DVV} in fact suggests that this is
an exact linearized equation.
}
\EQ
L_0(\mu)T \equiv [{1 \over 2}(\partial_t^2 -\partial_{\phi}^2)
-\sqrt{2}\partial_{\phi} + \mu\e^{-\sqrt{2}\phi}]T=T.
\label{viracon2}
\EN
Given this form of the tachyon equation, a very plausible resolution
of the puzzle is that the field $\zeta$ is related through a
non-local field redefinition of the following form
\EQ
T(t,\phi)=e^{-\sqrt{2}\phi}\int_0^{\infty}d\sigma
\exp (-2\sqrt{2\mu}\e^{-\phi/\sqrt{2}}\cosh\sigma) \gamma(i\partial_{t_0})
\partial_{\sigma}\zeta(t_0,\sigma)
\label{mstrans}
\EN
where $\gamma(i\partial_{t_0}))=\gamma(-i\partial_{t_0})^*$ is an
arbitrary weight function to be determined by the requirement of
normalization. It is easy to check that this is an intertwining
operator for the correspondence
\EQ
L_0(\mu)T \Leftrightarrow {1 \over 4}(\partial_{t_0}^2-
\partial_{\sigma}^2)\zeta,
\label{inttwi1}
\EN
provided $\partial_{\sigma}^2\zeta|_{\sigma =0}=0$ which is
guaranteed for the on-shell linearized solution satisfying the
Dirichret condition. This transformation was first pointed out
by Moore and Seiberg\cite{MS} in order to connect the macroscopic loop
operator to the collective field. For our later purpose,
it is more convenient to cast it into a Fourier transform form
\EQ
\tilde T(t,\phi)= \int_{-\infty}^{\infty}dp \tilde \zeta(p)
\gamma (p) K_{ip}(2\sqrt{\mu}\e^{-\phi/\sqrt{2}})\e^{-ipt_0}
\label{mstrans2}
\EN
where
\EQ
\zeta(t_0,\sigma)=\int_{-\infty}^{\infty}{dp \over p}\tilde \zeta (p)
\e^{-ipt_0}\sin p\sigma.
\label{zetasin}
\EN
The effect of this transformation on the S-matrix can be read off
from the asymptotic behavior for large $\phi$,
\EQ
\tilde T (t, \phi) \sim
\int dp \tilde \zeta(p) \gamma (p) \e^{-ipt_0}
(\Gamma(ip)\mu^{-ip/2}\e^{ip\phi/\sqrt{2}}
+\Gamma(-ip)\mu^{ip/2}\e^{-ip\phi/\sqrt{2}}).
\label{asympt1}
\EN

This implies that the on-shell wave functions are in general multiplicatively
renormalized by momentum-dependent
factor, $\gamma(p)\Gamma(\pm ip)\mu^{\mp ip/2}$
when we move from the collective
field $\zeta$ to the tachyon field $\tilde T$.
Requiring the unitarity of the S-matrix, however, the renormalization
factor must be at most a pure phase. Then a natural choice for $\gamma (p)$
would be  $\gamma(p)=\Gamma(\pm ip)^{-1}$.
The continuum calculation of the correlation functions indeed suggests that
an S-matrix element continued to the \underline{Eucliden} region
$p \rightarrow i|q|$ take
the form
\EQ
(\prod_{i=1}^N-\mu^{|q_i|/2}{\Gamma(-|q_i|) \over \Gamma (|q_i|)})
A_{{\rm coll}}(q_1,q_2, \ldots,q_N)
\label{contcorrel}
\EN
where $A_{{\rm coll}}(q_1,q_2,\ldots,q_N)$ is an  S-matrix element of the
collective field $\zeta$ continued to purely imaginary momenta and
energies. (See, e. g., refs. \cite{GK}\cite{DFK}).
Remarkably enough, the on-shell amplitude $A_{{\rm coll}}$ behaves
as polynomials\cite{DJR} with respect to
 the absolute values of certain combinations of energies
and hence does not exhibit any pole singularity even in the tree
approximation. On the other hand, the multiplicative factor
can be understood as the Euclidean continuation
($p \rightarrow \pm i|q|$) of the
renomalization factor appeared in (\ref{asympt1}),
$\mu^{\mp p/2}{\Gamma(\pm ip) \over \Gamma(\mp ip)}$.
It is improtant to note that although being as pure phases
these factors have no physical effect in the Minkowski metric
, they carry information of the  background tachyon
condensate. For the poles at $|q|=1,2,3,\ldots$, coming from the
numerator of the multiplicative factor can be, in terms of the
energy of the tachyon $p_t=in/\sqrt{2}, (n=1,2,3,\ldots)$,
 interpreted as the values of momenta at which one-particle
tachyon wave resonates with the background tachyon condensation.
If an  incoming one-particle tachyon wave  produces $N-1$ outgoing
tachyons, resonances are  expected to occur
when both the momentum and energy conservation laws hold,
\EQ
i\sqrt{2}p_t=-(r+N-2), \\\ N \ge 2, \\\ r=1,2,3,\ldots
\label{resocon1}
\EN
where $r$ is the number of  insertions of the operator
$\e^{-\sqrt{2}\phi}$ corresponding to tachyon  condensation
and $-2$ originates from the vacuum charge carried by
the $\phi$ coordinate.
{}From (\ref{contcorrel}) using the energy conservation
$|q_N|=\sum_{i=1}^{N-1}|q_i|$, we see that the residue of the
resonace pole at $|q_N|=r+N-2$ is proportional to
$(\prod_{i=1}^{N-1}R_L(q_i))A_{{\rm coll}}$ where
\EQ
R_L(q)=-\mu^{|q|}{\Gamma(-|q|) \over \Gamma(|q|)}
\label{refl1}
\EN
is nothing but the Euclidean-continued reflection coefficient for
the tachyon wave.

\subsection{black-hole background}

Let us next briefly summarize the relevant  properities of the
exact black-hole solution of Witten. The solution is described by an
$SL(2,R)/O(1,1)$ (or $SL(2,R)/U(1) $ in the case of Euclidean black hole)
gauged WZW model with $k=9/4$. In view of the role played by the linearized
tachyon equation in the usual $c=1$ case as discussed above,
we first focus on the Virasoro condition for tachyon fluctuation
which has been discussed in detail by Dijkgraaf, Verlinde and Verlinde
\cite{DVV}.

Let us parametrize the $SL(2,R)$ group manifold as
\EQ
g=\pmatrix{a  & u \cr
           -v & v\cr}  ,\quad uv+ab=1. \label{sl2para}
\EN
The parameters $a, b$ are redundant because of  gauge symmetry
$g \rightarrow hgh, \, h \in O(1,1)$. In terms of $u, v$, the
Virasoro condition takes the form
\EQ
L_0(u,v)T\equiv {1 \over k-2}[(1-uv)\partial_u \partial_v -
{1 \over 2}(u\partial_u + v\partial_v)-{1 \over 2k}
(u\partial_u-v\partial_v)^2]T=T.
\label{viracon3}
\EN
In fact, the Virasoro operator $L_0(u,v)$ consists of
two parts, $L_0= -\triangle_0 + (u\partial_u-v\partial_v)^2/4$,
where $\triangle_0$ is the Casimir operator of $SL(2,R)$.
The on-shell  tachyon corresponds to the continuous series
representation of $SL(2,R)$ which has eigenvalues
$\triangle_0= -\lambda^2 - {1 \over 4} \, \quad (\lambda = {\rm  real})$
and  $-i\partial_t= 2i\omega$ with the on-shell condition
$\lambda^2 = 9\omega^2$ at $k=9/4$.
When this equation is rewritten in a covariant form with background
spacetime metric $G_{\mu\nu}$ and dilaton $\Phi$,
$
L_0=-{1 \over  2e^{\Phi}\sqrt{G}}\partial^{\mu}e^{\Phi}
\sqrt{G}G_{\mu\nu}\partial_{\nu}
$,
we have ``exact''
(in the sense of $\alpha'$ expansion) expressions for the  background fields,
\EQ
ds^2 = {k-2 \over 2}[dr^2 - \beta^2(r)d\bar t^2],
\label{exactmetric}
\EN
\EQ
\Phi = \log (\sinh r/\beta(r)) + a,
\label{exactdil}
\EN
\EQ
\beta(r)= 2(\coth^2{r \over 2} -{2 \over k})^{-1/2},
\label{beta}
\EN
where the new coordinate $r$ and time $\bar t$ are defined by
\EQA
u &=& \sinh{r \over 2} \e^{\bar t},
\label{urt}\\
v &=& -\sinh{r \over 2} \e^{-\bar t}.
\EQN
The variables $(r, \bar t)$ are good coordinates for
describing the static exterior region outside
 event horizon sitting at $r=0$. The arbitrary constant $a$
 is related to the ADM mass of the black hole
by
\EQ
M_{{\rm bh}}= \sqrt{{2 \over k-2}}\,\e^a
\label{bhmass}
\EN
as shown in ref. \cite{Witten1}. The expressions
(\ref{exactmetric})$\sim$(\ref{beta})
reduce in the limit $k \rightarrow \infty$ to the solution of a
low energy approximation of the $\beta$-function condition. It should be
kept in mind, however, that in 2D critical string theory there
is no nontrivial systematic $\alpha'$-expansion, since the string
coupling always vanishes in that limit.

The global geometric structure
 described by the exact metric (\ref{exactmetric}) has
recently been shown to be free of curvature singularity and
consists of an infinite copies of a black-hole type spacetime
connected by wormholes at $uv=1$.  However, it should be
noted that there still is a ``dilaton singularity'' at $uv=1$
where the string coupling ${\rm g}_{{\rm st}}\sim \e^{-\Phi/2}$
diverges. In terms of $u, v$, $\Phi$ reads
\EQ
\Phi=\log[4\{-uv(1-uv)(-{(1-uv) \over uv}-{2 \over k})\}^{1/2}]+a.
\label{uvdila}
\EN
The region $uv >1$ defines a
disjoint region with a naked singularity.

The free parameter $a$ can be eliminated by making a scale transformation
\EQ
u \rightarrow \sqrt{M}^{-1}u, \, v \rightarrow \sqrt{M}^{-1}v,
\quad M\equiv \e^a.
\label{scale}
\EN
which in turn introduces the black-hole mass parameter in
a more familiar looking manner. Namely, at a price of eliminating
the  parameter $a$, we have to replace $1-uv$ by $M-uv$
in (\ref{viracon2}) and (\ref{uvdila}).

Here we emphasize a fact that since the string coupling is
determined by the background dilaton field, we have a simple
relation
\EQ
{\rm g}_{{\rm st}}(r=0)  \propto \e ^{-a/2}= M^{-1/2}
\label{couplmass}
\EN
This is to be contrasted with the dependence of
$g_{{\rm st}} \propto \mu^{-1}$
on the cosmological constant in flat spacetime.
This observation will play a crucial role in our proposal.
It also casts doubts on previous works\cite{Das,Dhar,Russo,Yang} on the
relation of black hole solution and matrix model.

Let us now consider the asymptotic  behavior of the Virasoro
condition and the dilaton field for $r\rightarrow \infty$ using
$u \sim  \e^{{r \over 2}+\bar t}, \, v \sim \e^{{r \over 2}-\bar t}$.
\EQ
L_0\sim {1 \over 4(k-2)}(\partial_r^2 + \partial_r)+
{1 \over 4k} \partial_{\bar t}^2,
\label{asympvira}
\EN
\EQ
\Phi\sim r + a - \log 4.
\label{asympdila}
\EN
Thus, the parameters $r, \bar t$ are identified asymptotically with the
$\phi$ and $t$ for the linear dilaton background as
\EQA
\bar t &\leftrightarrow& \sqrt{{1 \over 2k}}\, t = {\sqrt{2} \over 3}\,t,
\label{dic31}\\
r &\leftrightarrow& \sqrt{{2 \over k-2}}\,\phi = 2\sqrt{2}\,\phi.
\label{dic32}
\EQN
In terms of momentum and energy, the dictionary reads
\EQA
ip_{\phi} &=& -\sqrt{2} + i2\sqrt{2}\lambda = -\sqrt{2} +
{i \over \sqrt{2}}p_{\sigma},\label{dic41}\\
ip_t &=& i{2\sqrt{2} \over 3}\omega ={i \over \sqrt{2}} p_{t_0}. \label{dic42}
\EQN
We see that there is a one-to-one correspondence
of tachyon states between  the black-hole and linear
dilaton backgrounds.

The discrete states around the black-hole
background have been studied by several authors
\cite{DN,BK,EKY}. In the Minkowski metric, the spectrum of the
discrete states is isomorphic to that of the linear
dilaton background with just  the same values of
momenta and energies determined by the
condition for the discrete states for the latter through the
above dictionary. In particular,
the first nontrivial discrete state with zero energy
($j=1, m=0$ or $ip_{\phi}= -2\sqrt{2}, \, p_t =0$) is
identified with the operator associated with the mass
of black hole, as can be seen from  the first
correction to the asymptotic behavior of the exact spacetime
metric,
\EQ
ds^2 \sim {k-2 \over 2}[dr^2 - {4k \over k-2}(1-
{4k \over k-2}e^{-r} + {\rm O}(\e^{-2r}))dt^2].
\label{asympmetric}
\EN
It is crucial to note that the $\phi$ momentum
is twice that of the operator corresponding to
tachyon condensation. This explains, for example, the
difference in the relation between string coupling and
the black hole mass or the fermi energy of the matrix model,
$g_{{\rm st}}\sim M^{-1/2}$ versus $\mu^{-1}$.

Finally,  before going into  our main issue  we briefly
describe the properties of the solutions of the
Virasoro condition. The integral representation of the
solution with definite $\omega$ and $\lambda$ is
\EQ
\int_C {dx\over x} x^{-2i\omega}(\sqrt{M-uv}+{u \over x})^{-\nu_-}
(\sqrt{M-uv} -vx)^{-\nu_+}
\label{interep}
\EN
with $\nu_{\pm} = {1 \over 2}-i(\lambda \pm \omega)$. Following
ref. \cite{DVV}, we denote two independent solutions
corresponding to the coutours $C_2\equiv[-u\sqrt{M-uv}, 0],\,
C_4\equiv]-\infty,\nu^{-1}\sqrt{M-uv}] $ as ($y\equiv uv =-\sinh^2 {r \over
2}$)
\EQA
T_{C_2}&=& U_{\omega}^{\lambda}=\e^{-2i\omega \bar t}F_{\omega}^{\lambda}
(y),\label{tc2}\\
T_{C_4}&=& V_{\omega}^{\lambda}=\e^{-2i\omega \bar t}F_{-\omega}^{\lambda},
(y),\label{tc4}
\EQN
\EQ
F_{\omega}^{\lambda}(y)= (-y)^{-i\omega}B(\nu_+, \nu_-)
F(\nu_+, \nu_-. 1-2i\omega, y)\label{ffunction}.
\EN
The asymptotic behaviors of the solutions are, apart from an
irrelevant common phase factor,
for $r \rightarrow 0 \, ({\rm horizon})$:
\EQA
U_{\omega}^{\lambda} &\sim& \beta(\lambda, \omega)
({u \over \sqrt{M}})^{-2i \omega},
\label{uasymp}\\
V_{\omega}^{\lambda} &\sim& \beta(\lambda, -\omega)
(-{v \over \sqrt{M}})^{-2i \omega}
\label{vasymp}
\EQN
and for $r \rightarrow \infty$ (null infinities):
\EQ
F_{\omega}^{\lambda}\sim \alpha (\lambda, \omega)(-y)
^{-{1 \over 2}+ i\lambda}
+ \alpha (-\lambda,\omega)(-y)
^{-{1 \over 2}- i\lambda}\label{fasympt}
\EN
where
\EQA
\alpha(\lambda, \omega) &=& {\Gamma(\nu_+)\Gamma(\bar \nu_- - \nu_+)
\over \Gamma(\bar \nu_-)},\label{alphacoeff}\\
\beta(\lambda, \omega) &=& B(\nu_+, \bar \nu_-).
\label{betacoeff}
\EQN
{}From these behaviors, we  see  that $U_{\omega}^{\lambda}$ describes
a scattering of a wave coming from past null infinity with
the black hole, while $V_{\omega}^{\lambda}$ describes a wave emitted
by the white hole crossing the past event horizon. In the followings,
we will deal with the solution $U_{\omega}^{\lambda}$
since we are interested in the S-matrix elements of tachyons,
 incoming from the asymptotic flat region at part null infinity and
scattered out to future null infinity,
 which are the only legitimate observables
in string theory.
For this solution with  on-shell  condition $\omega = 3\lambda \,(>0)$,
ratios of the coefficients appearing in the above asymptotic behaviours,
\EQA
R_B(\lambda)&=& {\alpha(\lambda, \omega)\over
\alpha (-\lambda, \omega)},
\label{refleccoeff}\\
T_B(\lambda)&=& {\beta(\lambda, \omega)\over
\alpha (-\lambda, \omega)},
\label{absorbcoeff}
\EQN
are reflection and abosorbtion coefficients, respectively,
 for the  scattering of
 tachyon  wave with a static black hole,  and satisfy
the unitarity relation
\EQ
|R_B|^2 + {\omega \over \lambda} |T_B|^2 = 1
\label{unitarity}.
\EN
A prefactor  ${\omega \over \lambda}$ in the lhs  is necessary
to account for the change of velocities  of  tachyon
at the horizon and at the null infinity. This is caused by the
violation of equivalence principle due to the presence of
dilaton background.

\section{Working Hypothesis}
\setcounter{equation}{0}

What lessons do we have to learn from the review of the
spacetime interpretation of the $c=1$ matrix model in
attempting to extend that to black-hole background?
First it is clear
that there is no way of detecting nontrivial background
dependence by simply looking at the linearized field
equation of the scalar collective field $\zeta$. We have
seen that the linearized equation around an arbitrary
static solution of the matrix  model is reduced to
a free massless Klein-Gordon equation by choosing the
time-of-flight coordinate, irrespective of the form
of the hamiltonian of the matrix model. The  background
dependence of critical string theory, which is
nontrivial even in two dimensions because of the
coupling of dilaton, comes into the scene only after
making a non-local field redefinition. The
comparison as we made among the S-matrix  elements of
collective  field theory, the asymptotic behavior of
the redefined field, and the results on non-bulk
correlation functions  in the euclidean continuum
approach, strongly suggests that the momentum
dependent external leg factor of the continuum correlation
functions should be interpreted as an indication of the
necessity of a non-local  field redefinition for the collective
field $\zeta$. We have argued that although the external
leg factor is a pure phase in the Minkowski space and hence
is not observable, it encodes an important physical
information about the background as manifested as resonance
behavior at purely imaginary discrete values of momenta or energies.
We note  that this proposal is similar to that of Polyakov
\cite{Polyakov}
, but is in fact slightly different in that the residue
of the resonace pole need not necessarily equal to the
free field correlator.
A very remarkable point of this interpretation is that
the coefficient appearing in the solution of the single-particle
problem simultaneously accounts for the resonace poles for the
multiparticle amplitudes. As argued by Polyakov, we expect
that the codimension 2 pole singularities may only appear
in external legs. Presumably, the consistency of this
peculiar behavior of the on-shell amplitudes is
deeply related with the $W_{\infty}$ symmetry of the 2D
string theory and the matrix model.

Now in the case of the black hole background, no reliable
continuum calculation of the scattering ampltitudes
has been done yet. Naively, one may hope to apply the method
of free-field representation of the WZW model in
calculating the Euclidean correlation functions as in the
same way as being applied to the Liouville model coupled with
$c=1$ matter, making an analytic continuation with respect
to the number of insertions of the black-hole mass operator.
A free-field representation of the $SL(2,R)/U(1)$ model
has indeed been known from the work of Bershadski and Kutasov
\cite{BK}.
However, as far as we know, it is very difficult to reproduce
even the two-point function such as the reflection coefficient
$R_B$ in this approach. It seems that the perturbation with respect
to the black-hole mass is much more dangerous than that of
the tachyon condensation. See, however, ref. \cite{MV} for a different
possibility.

In view of this situation, we would like to propose an
ansatz as a working hypothesis that the above mentioned
properties of the scattering amplitudes
also extende to the case of a black-hole background.
Namely, {\it the hypothesis is that the S-matrix elements are
in general factorized products of external leg factors  and
the amplitudes of a collective field theory}.
The external leg factors must  be related  to the
asymptotic behavior of  solutions of the linearized
tachyon equation as above. On the other hand, the collective
field theory should represent an appropriate deformation
of  the usual $c=1$ matrix model at the extreme limit
$\mu \rightarrow 0$ (zero fermi energy)
since we are considering the
background without tachyon condensation.
We emphasize that we consider the nonzero
cosmological constant (fermi  energy $\mu$)
and the black hole mass $M$ to be
two entirely different deformations of the same
crtical theory described by linear dilaton background.
In this main point, we differ from some
of related works\cite{Das,Dhar,Russo}. For other related issues,
see also \cite{ellis}.

The deformation that we would like to consider
must properly describe the effect of finite black hole mass.
Since the exterior region of black-hole where the
scattering experiment is done can be
regarded as static, we are entitled to assume that the deformed
matrix model is time-translation invariant as in the usual case.
It is also reasonable to expect that the collective amplitudes are
essentially polynomials since all  the codimension 2 resonance
poles should be  only in the external leg factors.
A crucial property that we have to maintain is that the string
coupling is proportional to $M^{-1/2}$ with $M$ being the deformation
parameter of the deformed matrix model corresponding to the
black hole mass. We assume that the action of the deformed
model is analytic with repsect to $M$, since it appears analytically
in the Virasoro condition.

Let us now study the possible resonance poles for the black hole bakground.
{}From the asymptotic behavior (\ref{fasympt}) and the associated
reflection coefficient (\ref{refleccoeff}), we see that the positions
of the resonance poles are
\EQ
i4\lambda = i{4 \over 3} \omega = i\sqrt{2}\,p_t =-2, -4, -6, \ldots.
\label{resocon2}
\EN
This contrasts with the case of the usual $c=1$ model where we
have poles at all negative integers, (\ref{resocon1})
 of the corresponding energy.
On the other hand, if we consider an amplitude for
an incoming tachyon with producing $N-1$ outgoing
tachyons, the energy and momentum conservation laws
are satisfied when the incoming tachyon energy obeys
\EQ
i\sqrt{2}p_t = -(2r + N-2)
\label{resocon3}
\EN
where $r$ now counts the number of insertions of the
black hole mass operator. The factor $2$ multiplied to $r$
comes about because the momentum carried by the
black hole mass is twice that of the tachyon condensation.
Comparing (\ref{resocon2}) and (\ref{resocon3}), we see that
our hypothesis is consistent when only even $N$ are allowed.
This puts a stringent requirement to our hypothesis: the
on-shell scattering amplitudes of the
deformed matrix model must vanish when $N=$ odd. It is
clear that the usual $c=1$ matrix model with the purely
oscillatory potential can never
comform to this property.

Our task now is firstly to establish the possiblity of
a non-local field redefinition connecting the Virasoro condition
for the black hole background to the free massless Klein-Gordon
equation, and secondly to find a deformed matrix model satisfying
all of our criteria.

\section{Non-Local Field Redefinition}
\setcounter{equation}{0}

We first deal with  the non-local field redefinition.
Consider the integral representation (\ref{interep})
with the contour $C_2$ which is appropriate for
the scattering problem in the exterior region $(u>0, v<0)$.
\EQ
U_{\omega}^{\lambda}(u,v)=
\int_{C_2}{dx \over x} x^{-2i\omega}(\sqrt{M-uv}+ {u \over x})^{-\nu_-}
(\sqrt{M-uv}-vx)^{-\nu_+}.\label{uint}
\EN
Since the spectrum of the on-shell solution has a one-to-one
correspondence through (\ref{dic41}) and (\ref{dic42}) with
that of the free Klein-Gordon equation, it is natural to
make the following change of integration variable,
\EQA
(\sqrt{M-uv} + {u \over x})^{-1}(\sqrt{M-uv} - vx)
&=& \e ^{-4t_0 /3} \label{intch1},\\
(\sqrt{M-uv} + {u \over x})(\sqrt{M-uv} - vx)
&=& \e ^{-4\sigma} \label{intch2}.
\EQN
Then apart from an irrelevant numerical coefficient, the
solution takes the form
\EQ
U_{\omega}^{\lambda}
=\int_{-\infty}^{\infty}dt_0 \int_0^{\infty} d\sigma\,
\delta({u\e^{-2t_0/3}+v\e^{2t_0/3} \over 2} - \sqrt{M} \cosh 2\sigma)
\e^{-4i\omega t_0/3}\cos 4\lambda \sigma.
\label{interep2}
\EN

It is indeed easy to show that the $\delta$-function kernel here is an
intertwining operator between the Virasoro operator (\ref{viracon3})
and the free Klein-Gordon operator for $k=9/4$. We recognize the
form of  plane wave of the $\sigma$-derivative of
$\zeta$ satisfying the Dirichret boundary condition in the rhs.
We note that, according to the
dictionary (\ref{dic41}), (\ref{dic42}), (\ref{dic21}) and
(\ref{dic22}), the correspondence of
momentum  and energy is
\EQ
p_{\sigma}=4\lambda,\quad p_{t_0} = {4 \over  3}\omega.
\label{dic5}
\EN
Thus the non-local field redefinition now  is given by
\EQ
T(u,v) =
\int_{-\infty}^{\infty}dt_0 \int_0^{\infty} d\sigma\,
\delta({u\e^{-2t_0/3}+v\e^{2t_0/3} \over 2} - \sqrt{M} \cosh 2\sigma)
\gamma (i\partial_{t_0})\partial_{\sigma}\zeta (t_0, \sigma)
\label{bhfieldred}
\EN
where $\gamma(i\partial_{t_0})^* = \gamma (-i\partial_{t_0})$
is an arbitrary weight function to be fixed by normalization
condition,  as before.

In terms of the Fourier decomposition (\ref{zetasin}), it reads
\EQ
T(u,v) = \int_{\infty}^{-\infty}dp\,
\tilde  \zeta (p) \gamma (p) U_{\omega (p)}^{\lambda(p)}(u,v)
\label{fourier2}
\EN
with $\omega (p) = 3p/2, \lambda (p) = p/2$. In particular, the
asymptotic behavior for $y\rightarrow \infty$ is
\EQA
T(u,v) = \int_{-\infty}^{\infty}dp\,
\bar \zeta (p) \gamma (p)
&[&(-y)^{-{1 \over 2} + i \lambda (p)}\alpha (\lambda (p),\omega(p))
+\nonumber \\
&&
(-y)^{-{1 \over 2} - i \lambda (p)}\alpha (-\lambda (p),\omega(p))]
\e^{-2i\omega(p) t}.
\label{fourierasymp}
\EQN
This shows that an asymptotic wave packet of $\zeta$ field is
transformed into a deformed wave packet of the tachyon field.
Note that there is no singularity in the weight functions
for real $p$.

A remarkable and somewhat puzzling fact in this transformation is
that even if $\zeta$ is a plane wave satisfying a perfectly
reflecting boundary condition, the transformed wave has both
the reflected and absorbed parts. See (\ref{uasymp}) and (\ref{unitarity}).
After all, naturalness of the transformation can only be
judged when it is combined with the deformed matrix model.

We here remark that  similar transformations as (\ref{interep2}) have
been known for sometime. However, we must emphasize that our
interpretation is different from other works\cite{Martinec}
\cite{Das,Dhar}. Most importantly,
we do not identify the expression $\sqrt{M}\cosh 2\sigma$ with the
eigenvalue coordinate of the usual $c=1$ matrix model. It would
have implied $M \sim \mu$, which identifies the
black hole mass with fermi energy, and contradicts with the
relation between the string coupling and the black hole mass,
$g_{{\rm st}}(r=0) \sim M^{-1/2}$.
As we will see in the next section, the correct interpretation
is that $\sqrt{M}\cosh 2\sigma$ is nothing but the
square of the eigenvalue coordinate of a deformed matrix model.

\section{Deformed Matrix Model}
\setcounter{equation}{0}
\subsection{Deformed hamiltonian}

In the limit of vanishing black hole mass, the black hole
background reduces to the linear-dilaton vacuum.
This is a singlular limit in the sense that
the string coupling diverges, corresponding to the $c=1$ matrix
model with vanishing 2D cosmological constant $\mu = 0$,
or  zero fermi energy. Since, according to our
hypothesis, the deformation corresponding to non-vanishing black hole
mass cannot be described by the usual matrix model,
 we have to seek for other possible
deformations than the one induced by a change of the
fermi energy.  We therefore assume that the fermi energy is
kept exactly at zero, while the hamiltonian itself is modified,
\EQ
h(p,x) \rightarrow h_M (p, x) = {1 \over 2}(p^2 -  x^2)
+ M \delta h (p, x).
\label{defham}
\EN
We have here assumed that the deformation is
described by a term linear in $M$. If that would not work,
we would have to add higher order terms. Fortunately, however,
we will have a natural candidate with only the linear term,
satisfying our criteria.

The first requirement for $\delta h$ is a scaling property
to ensure that the string coupling is proportional to $M^{-1/2}$.
The general form of the collective hamiltonian (\ref{collham})
and the field shift around the classical ground state
solution imply that the string coupling in general is
proportional to $({dx \over d\sigma})^{-2}$. Thus, our
requirement is satisfied if the deformation operator
$\delta h$ has a scaling property
$\delta h(p, x) \rightarrow \rho^{-2}\delta h(p,x)$
under a global scale transformation $(p, x) \rightarrow
(\rho p, \rho x)$. This leads to
\EQ
\delta h(p,x) = {1 \over 2 x^2} f({p \over x})\label{defscale}.
\EN

To infer the form of an undetermined function $f({p \over x})$,
we invoke the following observation on the algebraic
property of the double scaled matrix model. As mentioned
in section 2, the usual $c=1$
hamiltonian $h= (p^2 - x^2)/2$ allows a set of
the eigenoperators $A_{j,m}$ satisfying  the
Poisson bracket relation
\EQ
\{h(p,x), A_{j,m}\}=2mA_{j,m}
\label{pb1}
\EN
and the $W_{\infty}$ algebra
\EQ
\{A_{j,m}, A_{j,'m'}\}=(mj' -m'j)A_{j+j'-1,m+m'}.
\label{pb2}
\EN
The origin of this  algebraic structure which is supposed to encode
the extended nature of strings can be traced back to the existence
of an $SL(2,R)$ algebra consisting of
\EQA
L_1&=& {1 \over 4}(p^2-x^2)= h(p,x)\label{l1}, \\
L_2&=& -{1 \over 4}(px + xp)\label{l2}, \\
L_3&=& {1 \over 4}(p^2+x^2)\label{l3}.
\EQN
The existence of a set of  eigenoperators satisfying the $W_{\infty}$
algebra is related to this $SL(2,R)$ structure by
\EQ
A_{j,m}=L_+^{{j+m \over 2}}L_-^{{j-m \over 2}},  \, (L_{\pm}=L_3 \pm L_2)
\label{al}
\EN
which close under the Poisson bracket since the Casimir invariant has
a fixed value
\EQ
L_1^2 + L_2^2 - L_3^2= {3\hbar \over 16}.\label{cas1}
\EN
(Here we include the Planck constant to indicate the effect of
operator ordering.)

Since the spectrum of the discrete states of the
black hole background is essentially the same
as the usual $c=1$ model in the Minkowski metric,
it is natural to require that the deformed model
should also share a similar algebraic structure.
Fortunately, this is satisfied if the simplest choice
$f=1$ is made.
Namely, the $SL(2,R)$ generators are now given by
\cite{AFF}
\EQA
L_1(M)&=& {1\over 2}h_M (p,x)
={1 \over 4}(p^2-x^2+{M \over x^2}),
\label{l1m}\\
L_2(M)&=& -{1 \over 4}(px + xp),
\label{l2m}\\
L_3(M)&=& {1 \over 4}(p^2 + x^2 + {M \over x^2}),
\label{l3m}
\EQN
which satisfy
\EQ
L_1(M)+L_2(M)^2-L_3(M)^2= -{M\over 2}+{3\hbar \over 16}.
\label{cas2}
\EN
We  note that because of  different constraint
for the Casimir invariant  the  algebra of eigenoperators is now
modifed in an $M$-dependent way. Although
further implications of this property, especially
an understanding of its connection to the ground ring structure
of the $SL(2,R)/O(1,1)$ WZW model, must be  left for future investigations,
we believe that this already is a strong  motivation for
adopting this particular model as a serious candidate for our
deformed matrix model.

\subsection{Properties of the deformed matrix  model}

Let us proceed to  study the poperties of the  defomed model:
\EQ
h_M(p,x) = {1 \over 2}(p^2 - x^2) + {M \over 2 x^2}.
\label{defham2}
\EN
We assume here that $M>0$ and will consider the case $M<0$ in the
final section, separately.
The solution of the classical equation of motion with
energy $\epsilon$ is
\EQ
x^2(t_0) =-\epsilon +  \sqrt{M + \epsilon^2} \cosh 2t_0.
\label{classol}
\EN
The ground state corresponding to zero fermi energy is obtained
by setting $\epsilon =0$ and replacing the time variable $t_0$
by the time-flight coordinate $\sigma$, $x^2=\sqrt{M}\cosh 2\sigma$.
We recognize that this is precisely the quantity which appeared in
the integral transformation (\ref{bhfieldred}) whose $\delta$-
function gives us a relation between the black hole
and the matrix model variables, $
x^2={u\e^{-2t_0/3}+v\e^{2t_0/3}\over 2}$. The coupling
function is now determined to be
\EQ
g(\sigma) \equiv {\sqrt{\pi} \over 12}({dx \over d\sigma})^{-2}
= {1 \over 48}\sqrt{{\pi \over M}}({1 \over \sinh^2\sigma }
+{1 \over \cosh^2 \sigma})
\label{couplf}
\EN
which explains the required relation with the black hole mass and
the asymptotic property for large $\sigma$.

We emphasize here that  we are taking a  double
scaling limit which is different from the usual  case. Namely, we fixed the
fermi energy  exactly at  zero and tuned the form of the
potential in a particular way. In terms of the original
matrix notation, the potential  is given by
$V(\Phi)=\Tr (-{1 \over 2}\Phi^2 + {\bar M \over 2\Phi^2})$.
Then the genus zero free energy in the limit of
vanishing scaling parameter, $\bar M \rightarrow 0$, behaves like
$F \sim {N^2 \over 8\pi \sqrt{2}}\bar M \log {\bar M \over \sqrt{2}}$.
The double scaling  limit is thus the limit $\bar M \rightarrow 0,
\, N \rightarrow \infty$
with $M \equiv N^2 \bar M$ being kept fixed. After  the
usual rescaling, $\,x\equiv \sqrt{N}\times$ eigenvalue of $\Phi$,
the system is reduced to the free fermion system with the one-body
potential $-{1 \over 2}x^2 + {M \over 2x^2}$.
Note that in the limit $M \rightarrow 0$ the potential
approaches to the usual inverted harmonic oscillator potential with
a repulsive $\delta$-function like singularity. Although the
critical point itself is the same as in the usual model,
apart from the $\delta$-function like infinite wall at the
origin, the above limit defines a different massive  continuum
theory.

It seems worthwhile to point out that the $\delta$-function
like potential at the origin in the limit $M \rightarrow 0$
conforms to Witten's discussion \cite{Witt2}of the ground ring structure of
the $c=1$ Liouville theory with {\it vanishing} cosmological
constant. According to his arguments on the connection of Liouville
theory coupled with $c=1$ matter  to
the  matrix  model, the negatively dressed (i. e., forbidden
in Seiberg criterion) $W_{\infty}$ operators
such as the one corresponding to
the black  hole mass are concentrated at the origin of the phase
space $(p,x)$. This is indeed the case in our deformed model
in the limit $M \rightarrow 0$, since the states are
affected only near at
the top
of the fermi sea by the $\delta$ function potential
in the semi-classical limit. It is an interesting
open question whether this similarity can be made more precise.
We  should also note that
this in turn indicates a danger in treating the black hole
mass parameter as a perturbation to the model  without
the  deformation term. It is impossible to  reproduce the
properties of our model with finite $M$ in any finite
order of perturbation with respect to the mass parameter $M$.

Once the deformed model is identified as above, it is tempting to
extend (\ref{bhfieldred}) to the
unshifted collected field $\rho(t_0, x)$
as
\EQ
T(u,v)= \int_{-\infty}^{\infty} dt_0
\int_0^{\infty}dx \,
\delta({u\e^{-2t_0/3}+v\e^{2t_0/3} \over 2} - x^2)
\gamma (i\partial_{t_0})\rho(t_0, x)
\label{extfieldred}
\EN
A possible divergence at $x^2 \rightarrow \infty$ is eliminated
by choosing $\gamma (0) =0$ and only other possible source
of singularity will be the other end point $x=0$. However,
due to the infinite {\it repulsive} wall at $x=0$, we are
guaranteed that $\rho (t_0, 0) =0$. Hence, we do  not
expect any singularity in exact theory for (\ref{extfieldred}). The
physical  meaning of this   is not, however,  clear to us at present.

Let us finally mention a few related previous works.
The potential with a $1/x^2$ term has been
discussed by Avan and one of the authors (A. J.)\cite{AJ2}
as an example of an integrable collective field theory
other than the standard inverted harmonic oscillator model.
There, the
$SL(2,R)$ structure has also been noticed.
We should also mention that Z. Yang\cite{Yang}
 has proposed that the black hole
background should be described by the potential
$-{1 \over 2}(x-{M \over x})^2$ at vanishing fermi energy.
Superficially this might look similar to the present proposal, but
is actually quite different. Firstly, Yang identified
the tachyon field directly with a shifted collective field
without any field redefinition and
used it to exhibit a black hole type metric. This cannot be justified
since the linearized equation is then always recasted in the
free Klein-Gordon form by a change of coordinate.
The  metric is only an artifact
of the choice  of coordinates and does not take the
crucial dilaton coupling into account.
In fact, because of two-dimensionality, the metric
is quite arbitrary for massless Klein-Gordon equation.
Secondly,
the identification of the black hole mass in his proposal
is in
contradiction  with the relation $g_{{\rm st}} \sim M^{-1/2}$.
Thirdly, the system is singular from the beginning because of
the  infinite attractive potential at the origin and  the
fermi energy being assumed exactly on top of the potential.
In our case, the system is completely well defined as long as
$M>0$. In the final setion, we will discuss the singular case $M<0$
as identified
with the background  with naked singularity.

\subsection{Tree level scattering of $\zeta$ quanta}

We  now proceed to study the scattering of $\zeta$ quanta and to see
whether the amplitudes vanish for odd number of particles.
We will stay within the tree approximation in the
present paper. The most convenient way of deriving the
scattering amplitude in the tree approximation is to use
the classical fermi liquid picture \cite{Pol2,MP}. In this
picture, the $\zeta$ field is connected the profile functions
$\alpha_{\pm}(t_0, x)$ satisfying the equation of motion
(\ref{alphaeq}). In the case of deformed model, the general
solution to this equation is given in the following
parametrized form containing an aribitrary function
$\epsilon (s)$ describing the deviation of the
fermi surface from its ground state form,
\EQA
x(t_0,  s) &=&[-\epsilon (s) + \sqrt{ M + \epsilon^2(s)} \cosh 2(s-t_0)]^{1/2},
\label{parasol1}\\
\alpha(t_0, s)&=& {1 \over x(t_0, s)}\sqrt{M + \epsilon^2(s)}\sinh2(s-t_0).
\label{parasol2}
\EQN
On the other hand, the asymptotic form, for large $x$, of the profile function
satisfies the following relation
\EQ
\alpha_{\pm}(t_0, \sigma)=\pm x(\sigma)(1-{\psi_{\pm}(t_0\pm \sigma)
\over x^2(\sigma)}) + O({1 \over x^2}).
\label{asympalpha}
\EN
The functions $\psi_{\pm}(t_0\pm \sigma)$
represent incoming and outgoing waves, respectively.
By comparing this definition with that of the $\zeta$ field,
we have the relation,
\EQ
(\partial_{t_0}\pm \partial_{\sigma})\zeta =
\pm {1 \over \sqrt{\pi}}\psi_{\pm}(t_0\pm \sigma)
\label{psizeta}
\EN
for $t_0 \rightarrow \mp \infty$.

The relation between $\psi_+$ and $\psi_-$ can be established
by  studying the time delay. Let the times at which a parametrized point $s$
is passed by the incoming and outgoing waves at a fixed value of
large $\sigma$ be $t_1 (\rightarrow -\infty)$ and
$t_2 (\rightarrow \infty)$, respectively.
{}From (\ref{parasol1}) we have then
\EQA
(M+\epsilon^2(s))^{1/4}\e^{s-t_1} &=& M^{1/4} e^{\sigma},
\label{past}\\
(M+\epsilon^2(s))^{1/4}\e^{t_2-s} &=& M^{1/4} e^{\sigma}.
\label{future}
\EQN
This implies
\EQ
t_1+\sigma = t_2 -\sigma + {1 \over 2}\log (1 + {\epsilon^2 (s) \over M})
,\label{pastfuture}
\EN
and hence
\EQ
\epsilon (s) = \psi_+(t_1 +\sigma) = \psi_-(t_2 -\sigma).
\label{inout}
\EN
Thus we get functional scattering equations connecting
the incoming and outgoing waves
\EQ
\psi_{\pm}(z)= \psi_{\mp}(z \mp {1 \over 2}\log
(1+ {1 \over M}\psi^2_{\pm}(z))).
\label{scatt}
\EN
The result is  similar to that of the usual $c=1$ model, but
manifests a crucial difference in that it is invariant under
the change of sign of $\psi_{\pm}\rightarrow -\psi_{\pm}$.
This clearly ensures that the particles participating in
the scattering is even.

The explicit power series soluton of (\ref{scatt})
is
\EQ
\psi_{\pm}(z) = \sum_{p=0}^{\infty}{M^{-p} \over p!(2p+1)}
{\Gamma(1\pm {1 \over 2}\partial_z) \over \Gamma(1-p \pm
{1 \over 2}\partial_z)}\psi^{2p+1}_{\mp}(z)
\label{powerscatt}
\EN
which shows that the amplitudes are essentially polynomial with respect
to the momenta without any singularity.

The scattering equation
can also be rewritten in terms of  energy momentum tensor
\EQ
T_{\pm\pm}(z) = {1 \over 2\pi} \psi^2_{\pm}(z),
\label{emtensor}
\EN
as
\EQ
\int dz \e^{i\omega z} T_{\pm\pm}(z)
={M \over 2\pi}\int dz \e^{i\omega z}
{1 \over 1 \pm {i\omega \over 2}}
[(1 + {2\pi \over M}T_{\mp\mp}(z))^{1\pm{i\omega \over 2}}-1].
\label{emtensorinout}
\EN
One can easily check that this defines a canonical
transformation by confirming that the Virasoro algebra
(in the level of Poisson bracket) is preserved by this transformation.
This  relation for  the energy momentum tensor is very
similar to the one obtained recently by Verlinde and Verlinde
\cite{VV} for the S-matrix of the
$N=24$ dilaton gravity. The main difference is that in ref.
\cite{VV}
the  relation holds for {\it integrals} of the
energy momentum tensor while here it  is a
relation for the energy-momentum tensor itself.
Of course, dilaton gravity and our deformed
matrix model describe  inequivalent physical systems.
But, the similarity is striking and suggestive and
might turn out to be of further significance.

\subsection{Scattering and representations of collective field theroy}

The reader may wonder how the above result that the
S-matrix elements are nonvanishing only for even number
of particles could be compatible with the collective
hamiltonian which has only a cubic interaction.
To check the consistecy, let us compute the
S-matrix elements directly from the collective hamiltonian.
This will be also useful for
getting more insights on the nature of  collective
field theory in general.

The 3-point vertex in the hamiltonian in the interaction
representation takes the form
\EQA
H_3(t_0)&=&{\sqrt{\pi}\over 12}\int_0^{\sigma}d\sigma
({dx \over d\sigma})^{-2}\{(\Pi_{\zeta}-\partial^{\sigma}\zeta)^3
-(\Pi_{\zeta}+\partial^{\sigma}\zeta)^3\}\nonumber\\
&=&-{1\over 12\pi}\int d^3k\, f(k_1+k_2+k_3)\e^{-i(k_1+k_2+k_3)t_0}
\alpha(k_1)\alpha(k_2)\alpha(k_3)
\label{h3}
\EQN
where
\EQ
f(k)=\int_{-\infty}^{\infty}d\sigma \, ({dx \over d\sigma})^{-2}
\e^{ik\sigma},
\label{fk}
\EN
and we have defined the creation and annhilation operators,  $\alpha(k)$
with $k<0$ and $k>0$, respectively,
by
\EQ
\Pi_{\zeta}\pm \partial_{\sigma}\zeta = \pm {1 \over \sqrt{\pi}}
\int_{-\infty}^{\infty} dk \e^{-ik(t_0\pm \sigma)} \alpha (k),
\label{creannhi}
\EN
\EQ
[\alpha(k_1), \alpha(k_2)]= k_1\,\delta(k_1+k_2).
\label{creannhicomm}
\EN
Thus the on-shell 3-point ampltitude ($k_i >0$)
\EQ
-<0| \alpha(k_1)\alpha(k_2)\, i\int dt_0\, H_3(t_0) \,\alpha(-k_3)|0>
=ik_1k_2k_3f(0)\,\delta (k_1+k_2+k_3)
\label{onshell3}
\EN
vanishes if $f(0)=0$. Using the classical solution $x(\sigma)
=M^{1/4}\cosh^{1/2}\sigma$, we find
\EQA
f(k)&=& {1\over 4\sqrt{M}}{\cal P}\int_{-\infty}^{\infty}d\sigma
({1\over \sinh^2\sigma} +{1 \over \cosh^2\sigma})\e^{ik\sigma}\\
\nonumber
&=&{1 \over 4\sqrt{M}}{\pi  k (\cosh{\pi k \over 2}-1)\over
\sinh{\pi k\over 2 }}
\label{fk2}
\EQN
which indeed leads to $f(0)=0$. Here we adopted the principal-value
prescription,
$
{\cal P}F(\sigma)\equiv (1/2)[F(\sigma +i\epsilon)+F(\sigma -i\epsilon)]
$
,to deal with the singularity associated with the
turning point in the semi-classical approximation to the matrix model.
This has been argued to be the correct procedure in ref.
\cite{DJR}.

Next let us consider the case of the 4-point  amplitude,
\EQA
A(k_1,k_2,k_3;k_4)&\equiv&
<0|\prod_{i=1}^3 \alpha(k_i)({-1\over 2})
\int dt_1 \int dt_2 {\rm T}(H_3(t_1)H_3(t_2))\alpha(-k_4)|0>
\nonumber\\
&=&{1\over 2\pi i}\delta(k_1+k_2+k_3-k_4)k_1k_2k_3k_4  \nonumber\\
&\times& (F(k_2+k_3)+F(k_3+k_1)+F(k_1+k_2))
\label{4point}
\EQN
where
\EQA
F(k)&=& {\cal P}\int_{-\infty}^{\infty}d\ell {k-\ell \over \ell}
f(\ell)^2
\label{cfk}\\
&=&-{\pi^2 \over 16M}\int_{-\infty}^{\infty}d\ell\, \ell^2
{(\cosh{\pi\ell \over 2}-1)\over \sinh^2{\pi \ell \over 2}}.
\label{cfk2}
\EQN
Here we used
$
{\cal P}{1 \over x \pm i\epsilon}= {\cal P}{1 \over x}\mp i\pi\delta(x)
$
and $f(0)=0$. This is again a singular integral. Following
\cite{DJR}, we use the
$\zeta$-function regularization which for (\ref{cfk2}) amounts
to subtract a quadratically divergent piece $\int d\ell \,\ell^2$.
We then find
\EQ
F(k) = {2\pi \over 3M} \label{cfk3}.
\EN
Hence,
\EQ
A(k_1,k_2,k_3;k_4)= {-i \over M}\delta (k_1+k_2+k_3-k_4)k_1k_2k_3k_4.
\label{ans4point}
\EN
This coincides with the result obtained from the scattering equation
(\ref{powerscatt}).

Let us next move to higher-point amplitudes.
Using the formula in
\cite{DJR}
and $f(0)=0$, a 5-point
amplitude corresponding to a Feynman diagram (out of
15 different diagrams) generally takes
the following form
\EQA
A(k_1,k_2,k_3;k_4,k_5) &\propto&
\int_{-\infty}^{\infty} d\ell \int_{-\infty}^{\infty}d\ell'
f(\ell)f(\ell')f(\ell +\ell')\nonumber\\
&+&(k_1+k_2)(k_4 +k_5)
\int_{-\infty}^{\infty} d\ell \int_{-\infty}^{\infty}d\ell'
{1 \over \ell \ell'}f(\ell)f(\ell')f(\ell +\ell')
\label{5point}
\EQN
where $k_1+k_2$ and $k_4+k_5$ are the momenta of associated with
two internal lines of a Feynman diagram (note this uniquely fixes
the Feynman diagram).
These are again singular integrals. They can be
most conveniently treated by returning to the coordinate space
and using the principal-value prescription.
For example, the first term is then given as
\EQ
(2\pi)^2{\cal P}\int_{-\infty}^{\infty}d\sigma ({1 \over \alpha_0^2(\sigma)})^3
={(2\pi)^2 \over \sqrt{M}^3}{\cal P}\int_{-\infty}^{\infty}d\sigma
{(\cosh 2\sigma)^3 \over (\sinh
2\sigma )^6}
\EN
By making a change of integration variable $y=\sinh 2\sigma$,
we find the above integral is proportional to
$
{\cal P}\int_{-\infty}^{\infty}dy (y^{-6}+y^{-4})=0.
$
The second term is also rewritten in the coordinate space
\EQ
(2\pi)^2\int_{-\infty}^{\infty}d\sigma {1 \over \alpha_0(\sigma)^2}
(\partial^{-1}_{\sigma} {1 \over \alpha_0(\sigma)^2})^2
\propto
{\cal P}\int_{-\infty}^{\infty}d\sigma ({1 \over \sinh^2\sigma}
+{1 \over \cosh^2\sigma})(-\coth\sigma + \tanh  \sigma)^2
\EN
which again allows a change of variables so that it vanishes.
In general, one can see that this happens for all
odd-point amplitudes. Typically, one will have the
integral ($n=$positive integers)
\EQ
{\cal P}\int_{-\infty}^{\infty}d\sigma ({1 \over \alpha_0(\sigma)})^{2n+1}
\propto
{\cal P}\int_{-\infty}^{\infty}dy{(1+y)^n \over y^{4n +2}}
\EN
which all vanish by the principal value prescription.

However, we cannot deny the impression that the above calculations
are rather awakward in using {\it ad hoc} prescriptions for dealing
with singular integrals. Actually, the identical results can be
obtained without singular integrals, if we use the dual
representation in constructing the collective hamiltonian.
Namely, we can intoduce the collective field in a represention
in which the momentum $p$ is diagonalized instead of the
coordinate $x$. We introduce
the profile function $\beta(p)$ of the fermi sea as a function of $p$.
Denoting the two branches (corresponding to $x<0$ and $x>0$)
 of the fermi sea as $\beta_{\pm}$, the hamiltonian is now
\EQ
H= {1 \over 2\pi}\int dp \int_{\beta_-(p)}^{\beta_+(p)}dx
h(p,x)\equiv H_+-H_-, \label{pham1}
\EN
\EQ
H_{\pm}={1 \over 2\pi}\int_{-\infty}^{\infty} dp
[{1 \over 2}p^2\beta_{\pm}- {1\over 6}\beta_{\pm}^3-
{M \over 2\beta_{\pm}}].
\label{pham2}
\EN
Here we droped the boundaries  at $x\rightarrow \pm \infty$
because we are only interested in the dynamics of the
fluctuation of the fermi sea and the boundaries do not
participate in it. The Poisson bracket for $\beta_{\pm}$ and the
equation of motion are, respectively,
\EQ
\{\beta_{\pm}(p_1), \beta_{\pm}(p_2)\}= \pm 2\pi_1 \partial_p
\delta (p_1-p_2),
\label{poissonbeta}
\EN
\EQ
\dot \beta_{\pm}(t_0,p)=\partial_p h(p,\beta_{\pm}(t_0,p))
\label{betaeqmo}
\EN
The classical ground state solutiion is given as
\EQ
\beta_{\pm}=\mp[{p^2 + \sqrt{p^4+4M} \over 2} ]^{1/2}\equiv \mp\beta_0.
\label{betaground}
\EN
To study the fluctuation, an appropriate shift of
the field is $\beta_{\mp} = \mp \beta_0-\sqrt{\pi}({dp \over d\sigma})^{-1}
(\Pi_{\eta}\pm \partial_{\sigma}\eta)$ where
the relation with the time-of-flight coordinate and the
$p$-coordinate is
\EQ
{dp \over d\sigma}=\beta_0(p)+{M\over \beta_0^3(p)}\equiv \tilde \beta_0(p)
\label{psigma}
\EN
which is solved as $p={M^{1/4}\sinh 2\sigma \over \cosh^{1/2} 2\sigma}$
coinciding with ${dx \over d\sigma}$. Then the hamiltonian becomes
discarding the c-number part,
\EQA
H=\int_{-\infty}^{\infty}&d\sigma&
[{1\over 2}(\Pi_{\eta}^2+(\partial_{\sigma}\eta)^2)-
{\sqrt{\pi}\over 2}({1\over 6}-{1\over 2\beta_0^4}){1\over \beta_0^2(p)}
\{(\Pi_{\eta}+\partial_{\sigma}\eta)^3-(\Pi_{\eta}-\partial_{\sigma}\eta)^3\}
\nonumber \\
&-&{M \over 4\pi}\sum_{n=4}^{\infty}
(-)^n{\pi^{n/2} \over \beta_0^{n+1}(p)\tilde \beta_0^{n-1}(p)}
\{(\Pi_{\eta}+\partial_{\sigma}\eta)^n-(\Pi_{\eta}-\partial_{\sigma}\eta)^n\}].
\label{pham3}
\EQN
Note that we now have non-polynomial interactions, in contrast to
the purely cubic interactions of the collective field theory
in the $x$-representation.
Since in the $p$-representation there
is no mixing of left-right moving modes, we only retain
the right-moving terms in the interaction terms,
containing
\EQ
\Pi_{\eta} -\partial_{\sigma}\eta =-
{1 \over \sqrt{\pi}}\int_{-\infty}^{\infty} dk
\alpha(k)\e^{-ik(t_0-\sigma)}.
\label{right}
\EN
After substituting the expressions for $\beta_0, \tilde \beta_0$,
we get the follwing exlicit forms for the 3- and 4- point functions,
\EQ
H_3(t_0)={M^{-1/2}\over 12\pi}\int\,d^3k\e^{-i(k_1+k_2+k_3)t_0}
g_3(k_1+k_2+k_3)\alpha(k_1)\alpha(k_2)\alpha(k_3),
\label{ph3}
\EN
\EQ
g_3(k) =\int_{-\infty}^{\infty} d\sigma\, \e^{ik\sigma}
{\cosh 2\sigma (\cosh^2 2\sigma -3) \over
(\cosh^2 2\sigma + 1)}=-{\pi k \over 4}
{\sin ks_0 \over \cosh {\pi k \over 4}},
\label{ph31}
\EN
\EQ
H_4(t_0)={1 \over 4\pi M}\int d^4k \e^{-i(k_1+k_2+k_3+k_4)t_0}
g_4(k_1+k_2+k_3+k_4),
\label{ph4}
\EN
\EQA
g_4(k)&=&\int_{-\infty}^{\infty}d\sigma\, \e^{ik\sigma}
{\cosh^2 2\sigma \over (\cosh^2 2\sigma +1)^3}
\nonumber\\
&=&{(k^2+16)\sqrt{2}\pi \over 256}{\sin ks_0 \over \sinh {\pi k \over 4}}
+{\pi \sqrt{2} \over 512 \sinh {\pi k \over 4}}
(2\sqrt{2}k\cos ks_0 -12\sin ks_0),
\label{ph41}
\EQN
where $s_0 (>0)$ is a solution of $\cosh 4s_0 = 3$.
We see that all the integrals for the form factors are perfectly well
defined. Using these results, we find again that the on-shell
3-point ampltitude vanishes $g_3(0)=0$. The 4-point ampltitude now receives
the contribution from the 4-point vertex, in addition to
the one coming from a product of the 3-point functions.
They are, respectively,
\EQ
{-i\over 2M}\delta(k_1+k_2+k_3-k_4) k_1k_2k_3k_4\,g_4(0)\times 4!
\label{4pointcont}
\EN
and
\EQ
{-i \over 2M}\delta(k_1+k_2+k_3-k_4)k_1k_2k_3k_4\times
{3 \over 16}{\cal P}\int_{-\infty}^{\infty}
d\ell\, {k_1+k_2+k_3-\ell \over \ell}g_3(\ell)^2.
\label{3pointcont}
\EN
Evaluating the integrals, we find that the
coefficients after the products of momenta $k_i's$ are
$(3+15\sqrt{2}s_0)/4$ and $(5-15\sqrt{2}s_0)/4$, respectively, which
just reproduce the result (\ref{ans4point}).

We  expect that similar cancellations ocuring here must happen
in higher order terms as well, though checking this
explicitly in the Feynman graph
expansion becomes increasing tedious in the $p$-representation.
It is not difficult, however,  in general to convince ourselves
the equivalence with the method of Polchinski, using
the well  known general theorem that the computing
S-matrix elements in the tree approximation is equivalent
to obtaining a general solution to the classical equation
of motion.

It is remarkable that two entirely different field theories
give equivalent S-matrix elements.  The fields
$\zeta$ and $\eta$ can never be conneced by a simple field
redefinition. Remember that we have effectively interchanged
the role of field and coordinate, although  in the free fermion
picture, this is  simply a standard dual transformation interchanging
$x$ and $p$. This explains why we have the same
S-matrix elements in the usual $c=1$ model for both
cases of positive and negative cosmological constant
(see the first reference of \cite{Klebanov}).
 It might be useful to study
other possible choice of the representation such that
it manifests the vanishing of the S-matrix elements for
odd number of particles. Also, it is a very challenging question whether
a new symmetry of this kind can be generalized to higher dimesional
solutions of critical string theory.

\section{Discussions}
\setcounter{equation}{0}
\subsection{S-matrix elements for tachyon black-hole scattering}

Now we can proceed to discuss a plausible candidate for the
S-matrix elements
for tachyon black-hole scattering. Our basic working hypothesis
requires us to apply the non-local integral
transformation to each external line of a Green's function
of the collective field quanta $\zeta$ (or $\eta$) of the
deformed matrix model. Let the Green's function of the
$\zeta$ in the momentum representation $(k= p_{\sigma}=4\lambda,
E=p_{t_0}={4 \over 3}\omega)$ be
\EQ
G(k_1,E_1, k_2,E_2, \ldots, k_n,E_n)
\sim (\prod_{i=1}^n{1 \over k_i^2-E_i^2+i\epsilon})
A(k_1,k_2,\ldots,k_n)
\label{zetagreen}
\EN
Then, the asymptotic behavior of the trasnformed Green's function
for the tachyon field $\tilde T$ is given by
\EQA
\{\prod_{i=1}^n\int_0^{\infty}dk_i\int_{-\infty}^{\infty}dE_i
\gamma(E_i)\lambda_i\e^{-2i\omega_it_i}
(\alpha(-\lambda_i,\omega_i)\e^{-2i\lambda_ir_i}
&+&\alpha(\lambda_i,\omega_i)e^{2i\lambda_ir_i})\nonumber \\
\times {1\over k_i^2-E_i^2+i\epsilon}\}A(k_1,k_2,\ldots,k_n)
\label{tgreen}
\EQN
The integrals over $E_i's$ pick up the pole at $E_i= k_i$ or $-k_i$,
depending on $t_i>0$ or $t_i<0$, respectively. This implies that in the
asymptotic limit $r_i\rightarrow \infty$, the wave packets of
the tachyon field $\tilde T$ receive contribution from the
first  term. In other words, the incoming or outgoing wave
packets is right moving or left moving
respectively as it should be. If we choose a natural
normalization such that the incoming waves are  normalized to be unity,
$|\gamma(E_i)\lambda_i\alpha(-\lambda_i, 3\lambda_i)|=1$, the
outgoing waves have the normalization
\EQ
|\gamma(-E_i)\lambda_i\alpha(-\lambda_i, -3\lambda_i)|
=\bigl|{\alpha(-\lambda_i,-3\lambda_i\over \alpha(-\lambda_i,3\lambda_i)}\bigr|
=|R_B(\lambda_i)|.
\label{rbstar}
\EN
Remember the definition (\ref{refleccoeff})
of the reflection coefficient for the
black hole.
Thus, apart from pure phase factors, the outgoing external
lines get multiplied the reflection coefficient.

In the static black-hole  background geometry, part of the incoming
wave from the past null infinity is absorbed into the future horizon.
To have the complete set of the scattered waves, we therefore
need to take the absorbed waves into account.
The asymptotic behavior (\ref{uasymp}) shows that we have to
multiply the absorbtion coefficient corresponding to that part.

The rules for these external leg factors can be most
conveniently stated by using the set of creation and
annihilation operators for the asymptotic states $t\rightarrow \pm \infty$.
Let the asymptotic in and out oscillator of the $\zeta$ field
be $\alpha_{\pm}(\lambda)$ as before and
those of the $\tilde T$ fields on
past infinity (${\cal I}^+$), future infinity (${\cal I}^-$), and
on future event horizon (${\cal H}_+$), be
$\alpha_+^{{\cal I}}$, $\alpha_-^{{\cal I}}$ and
$\alpha_-^{{\cal H}}$, respectively. By convention,
the operators with $\lambda >0 (<0)$ are annhilation (creation)
operators such that
$[\alpha(\lambda_1), \alpha(\lambda_2)]=\lambda_1\delta(\lambda_1+\lambda_2)$
for each asymptotic region.
Suppose the scattering operator of the $\zeta$ fields is given by
\EQ
S_{\zeta}=\sum_{n,m=0}^{\infty}{1\over n!m!}
\prod\int_0^{\infty}\prod d\lambda_i
A(\lambda_1,\ldots,\lambda_m;\lambda_{m+1},\ldots,\lambda_{m+n})
\prod_{i=1}^m\alpha_-(\lambda_i)\prod_{j=m+1}^{n+m}\alpha_+(-\lambda_j).
\label{szeta}
\EN
Then, the scattering operator of the tachyon field $\tilde T$
is
\EQA
S_{\tilde T}&=&\sum_{n,m=0}^{\infty}{1\over n!m!}
(\prod\int_0^{\infty}\prod d\lambda_i)
A(\lambda_1,\ldots,\lambda_m;\lambda_{m+1},\ldots,\lambda_{m+n})\times
\nonumber \\
&&\prod_{i=1}^m(|R_B(\lambda_i)|\alpha_-^{{\cal I}}(\lambda_i)
+\sqrt{{\omega_i \over \lambda_i}}|T_B(\lambda_i)|\alpha_-^{{\cal  H}})
\prod_{j=m+1}^{n+m}\alpha_+^{{\cal I}}(-\lambda_j)
\label{tzeta}
\EQN
The unitarity of $S_{\tilde T}$, providing  the ``observations'' for
$t \rightarrow +\infty$ are made on both ${\cal I}^-$ and ${\cal H}^-$,
is ensured by the relation (\ref{unitarity}) and the unitarity of
$S_{\zeta}$.

It is natural to define a density operator for the out
states on ${\cal I}^-$ given an initial state $|i>^{{\cal I}}_+$ by
$\rho_S = \Tr_{{\cal H}^-}[S_{\tilde T}|i>^{{\cal I}}_+
<i|S_{\tilde T}^{\dagger}]$ where the trace is taken only over the
set of states on ${\cal H}^-$.

We would like,  however, warn the reader that although the above
S-matrix seems natural at least in the perturbative regime where we
just consider scatterings in a static black-hole background
by assuming the mass of black hole is sufficiently large, it is
by no means clear how to interpret it in the nonperturbative regime,
where we have to properly take into account the back reaction
of the geometry itself. Our tentative identification of the black hole mass
operator in the deformed matrix model
seems to indicate the impossibility of changing the black hole mass
by a dynmaical process. This seems consistent with an analysis
\cite{Russo2} of this question
using a low-energy effective field theory.
 Still,we have in general non zero
absorbtion coefficient.
Note, however, that the absorbtion
coefficient vanishes in the low energy limit.
Here it might be
worthwhile to remind ourselves that the relation of the
string coupling with the inverse black hole mass is analogous
to the situation in the soliton solutions of local field theories.
In the case of soliton solutions, the perturbation theory in fact
reproduces  the effect of back reaction order by order.
It is a very important question whether or not the
pertubation expansion in the present case can also be
interpreted as containing the effect of back reaction.
Saying anything about possible nonperturbative meaning
of our results would be dangerous without a more clear
understanding on this question.

\subsection{Naked singularity}

So far, we have been treating only the case $M>0$. When $M<0$,
the deformed matrix model has a singularity at the origin
corresponding to an infinite attractive force. Correspondingly,
the Virasoro condition contain a naked singularity at $uv=M$.
Equivalently, by making a transformation $u\rightarrow \sqrt{|M|}u,
\, v \rightarrow -\sqrt{|M|}v$, we can consider the same
Virasoro condition (\ref{viracon3}) in the region $uv>1,\, u>0, v>0$.
In terms of the exact global geometry described by (\ref{exactmetric}),
this region is disconnected from the exterior region $uv<0$.
In this subsection, let us work in the latter choice of the
coordinates.

Because of the singularity, a generic solution of the
Virasoro condition exhibits a logarithmic singularity
$\log (1-uv)$ near the naked singularity. However, it is
known that there is a particular solution which is regular at the
naked singularity, given by, $(y=uv)$
\EQA
W_{\omega}^{\lambda}&=&
\e^{-2i\omega t}\Gamma(\bar \nu_-)\Gamma(\nu_-)y^{-i\omega}
F(\nu_+, \bar \nu_-;1;1-y)
\label{w1}\\
&=&\e^{-2i\omega_it}
[{\Gamma(2i\omega)\Gamma(\bar \nu_-)\over \Gamma(\bar \nu_+)}
y^{-i\omega}F(\nu_+,\bar \nu_-;1-2i\omega;y)\nonumber\\
&+&{\Gamma(-2i\omega)\Gamma( \nu_-)\over \Gamma(\nu_+)}
y^{i\omega}F(\bar \nu_+,\nu_-;1+2i\omega;y)]
\label{w2}
\EQN
Using an  itegaral represention for $W_{\omega}^{\lambda}$,
\EQ
W_{\omega}^{\lambda}=e^{-2i\omega t}
\int_0^1 dz\, z^{-{1\over 2}+i(\lambda -\omega)}
(1-z)^{-{1\over 2}-i(\lambda -\omega)}(1-(1-y)z)^{-{1\over 2}+i(\lambda
+\omega)}
\label{wint}
\EN
and making a change of variable
\EQA
\e^{{2t_0 \over 3}}&=& ({1-z \over z})^{1/2}(1-(1-y)z)^{1/2},\\
\label{chv3}
\e^{2\sigma}&=& ({1-z \over z})^{-1/2}(1-(1-y)z)^{1/2},
\label{chv4}
\EQN
we can find an intertwining representation which is similar
to the black hole case. (A numercial proportionality constant is again
neglected.)
\EQ
W_{\omega}^{\lambda}=\int_{-\infty}^{\infty}d\sigma
\int_{-\infty}^{\infty}dt_0 \delta({u\e^{-2t_0/3}-v\e^{2t_0/3}
\over 2}-\sinh 2\sigma)
\e^{-4i\omega t_0 /3}\e^{-4i\lambda \sigma}
\label{wint2}
\EN
The $\delta$-function kernel in this result  is a natural generalization of
that  appeared in (\ref{interep2}). First, the minus sign in front of $v$ is
coming from the above trasnformation $v\rightarrow -v$. Second, in terms
of the time-of-flight coordinate, the classical ground solution
for $M<0$ case is
\EQA
x^2&=&\sqrt{|M|}|\sinh 2\sigma|,
\label{negaclass1}\\
p^2&=&\sqrt{|M|}|\sinh 2\sigma + {1 \over \sinh 2\sigma}|.
\label{negaclass2}
\EQN
Thus, after returning to our scaled coordinates,
the kernel takes the form
$
\delta({u\e^{-2t_0/3}+v\e^{2t_0/3}
\over 2}-\epsilon(x)x^2)
$
which is equivalent with the previous case  $M>0$, since in that case
we can restrinct the range $x$ to the positive real axis because of the
infinite repulsive potential at the origin.
Thus the difference is that the integartion range of $x$ or $\sigma$ is now
$(-\infty, \infty)$ instead of  the half infinite. Classically,
an incoming particle with total energy larger than a finite
critical value $\epsilon_0=-\sqrt{2|M|} (<0)$ reaches to the singularity in a
finite time. The extension to negative $\sigma$ may be interpreted
as a particle continues to move beyond the singularity.
It then seems natural to suppose that the tachyon wave which is chosen
to be regular at the naked singularity corresponds
classical fermi liquid passing through the singularity
of  the potential in the above sense.
It is then easy to apply again the fermi-liquid picture  to obtain the
scattering equation for the naked singularity.

Before doing that, let us briefly mention the asymptotic property
of the solution. By examining the asymptotic form for large $y$,
we easily see that  the waves are perfectly reflected
with the pure phase reflection coefficient
\EQ
R_N(\lambda)=-2^{-8i\lambda}
{\Gamma(1+4i\lambda)\Gamma({1\over 2}-4i\lambda) \over
\Gamma(1-4i\lambda)\Gamma({1\over 2}+4i\lambda)}
\label{nakref}
\EN
This exhibits the resonance poles at all integer values of
purely imaginary momenta $-4i\lambda =1,2,\ldots$, in sharp
contrast with the black-hole case with event horizon.
This difference is very mysterious to us. It seems as if
the special choice of perfectly reflecting boundary
condition effectively shifts the momentum of the mass operator
such that the resonance poles appear at the same values as
in the case of the usual
tachyon condensation, although the reflection coefficient is
not the same.
Whatever the interpretation of this peculior behavior
 might be, now there seems to be  no
reason for expecting vanishing of the odd-point on-shell
ampltitudes.

Let us derive the scattering equation for this case based on the
picture above. It is convenient to work  with $x^2$ and $p^2$.
For finite energy $\epsilon >\epsilon_0$, the solution of the
classical equation of motion is given in the parametrized form by
\EQA
x^2(t,s)&=&-\epsilon(s)+\sqrt{2|M|-\epsilon^2(s)}\sinh 2(s-t),\quad (t<\tau(s))
\label{negx1}\\
&=&
\epsilon(s)-\sqrt{2|M|-\epsilon^2(s)}\sinh 2(t+s-\tau),\quad (t>\tau (s))
\label{negx2}
\EQN
\EQ
\alpha^2(t,s)=x^2(t,s)-{1 \over x^2(t,s)}+2\epsilon (s)
\EN
where $\tau$ is determined by $\sinh 2(s-\tau)={\epsilon \over \sqrt{2|M|-
\epsilon^2}}$.
By comparing the asymptotic relation among the coordinate $\sigma$ and the
times $t_1, t_2$ as before at which a particular prametrized point $s$ of the
incoming and outgoing waves reach, we arrive at the relation
\EQ
t_1+\sigma =t_2-\sigma +{1\over 2}\log(1-{\epsilon^2(s) \over 2|M|})
-\sinh^{-1}{\epsilon(s)\over \sqrt{2|M|-\epsilon^2(s)}}.
\EN
This  shows that the scattering equation is
\EQ
\psi_{\pm}(z)=\psi_{\mp}
(z\mp {1\over 2}\log(1-{\psi_{\mp}^2(z) \over 2|M|})
\pm\sinh^{-1}{\psi_{\mp}(z)\over \sqrt{2|M|-\psi_{\mp}^2(z)}}).
\EN
The on-shell  ampltitudes are indeed non-vanishing for both odd
and even numbers of  particles. It is easy to check  that
the collective field theory in the $x$-representation has
a well  defined 3-point vertex which does not vanish on shell.

Although we are not so sure whether the difference in the properties of
tachyon scattering for $M>0$ and $M<0$ is an evidence for our
hypothesis, it is interesting that the deformed model discriminates
the two cases very clearly, as should be expected
from the exact linearized analysis of ref. \cite{DVV}.

\subsection{Conclusion}

We believe  that we have clarified some of the most crucial
steps in the problem of a 2D black hole background in the matrix-model
approach to string theory.
We have proposed a working hypothesis, based  on which we could
obtain several  concrete results. However, it is also clear
that we have left some of the crucial questions unanswered. We
have already emphasized a few of them. To conclude the
present paper, we will give  a list of most important remaining
problems.
\begin{enumerate}
\item  {\it Physical picture}: One might have expected that if the black
hole geometry is describable at all  in the matrix model, it had to be
related to the existence of the other side of the inverted harmonic
oscillator potential. In fact, however, after all our
experimentation, we have been forced to reach at a totally
different picture that the black hole should rather correspond
to a deformed matrix model with infinite repulsive wall
at the origin, such that the other side of the potential
becomes irrelevant. The effect of a horizon is mostly
taken  into  account by the non-local field redefinition, and  not  by
the form  of the potential.  It is
desirable to have some intuitive explanation why this must be so
if  our conclusion is ever correct.
\item {\it Unified understanding of the field redefinition and
the matrix model hamiltonian}: The question is whether and how
the matrix models themselves encode the information
of non-local field redefinition which is already background
dependent. We have seen that the characteristic properties of the
deformed matrix model, in particular,  its scattering ampltitudes,
are surprisingly consistent with the properties of the
one-particle wave functions coming from the
field redefinition. However, the coincidence looks rather accidental in
view  of lack of some more logical derivation.
We have no rigorous uniquness
proof for the field redefinition nor for the deformed model and
a unified derivation would illuminate both. In this unified theory, the
factorized ansatz for the S-matrix might be only a first
approximation.
\item {\it Tachyon condensation}:
To gain some insight into the above question, it might be
useful to study the effect of tachyon condensation in the
black hole background. In the present work, we have assumed that
the fermi energy is precisely zero. We conjecture that the
deformed matrix model with non zero  fermi energy corresponds to a black
hole  background with tachyon condensation.
In our approach, we should be able to generalize the non-local
field redefiniton so that it incorporates both (\ref{mstrans})
and (\ref{bhfieldred}) depending upon the
manitude of $M$ and $\mu^2$. As in the low energy
effective field theory\cite{Witten1,PPT},
the tachyon condensation provides an independent
deformation parameter of the  theory.
\item {\it Role of  $SL(2,R)$ algebra and the discrete states}:
We have mentioned the presence of an $SL(2,R)$ algebraic structure
in the deformed model as one of the motivations to our
proposal. Undoubtedly, that must be crucial for understanding
the full  symmetry of the theory. We have also to mention
the relevance of it for the question of possible $W_{\infty}$ hair of the
black hole \cite{ellis}.
Concerning to the discrete states, we do not have an adequate
understanding on the connection between the spectrum of
the discrete states and the poles in the external leg factors
in the case of the Minkowskian black hole. In the usual  $c=1$ case,
they are intimately related.
\item {\it Connection with the Euclidean black hole}:
We have only discussed the Minkowskian black hole. The significance
of our results for the Euclidean black hole is not
clear to us. If the integral transformation is formally
Wick-rotated, the periodicity in imaginary $t_0$ is $3\pi$.
The latter is to be identified by a standard argument with the
inverse Hawking temperature. To be consistent, the deformed
model should be the Euclidean version with that periodicity.
The question then arises whether our S-matrix is related
to Hawking  radiation or not. It might be useful to
remind that  when the information on the future event
horizon is summed up, our S-matrix operator turns into an
density operator as discussed  in a previous subsection. Is it
possible that the density operator obtained in this way is
related with the thermal density operator when
the contribution of the one-loop effect is  taken into account?
\end{enumerate}

We hope to return to some of these questions in the future.


\newpage
\noindent
Acknowledgements

\vspace{0.5cm}

We would like  to thank S. Das, I. Klebanov, D. Kutasov,
M. Li, P. Mende  and S.  Shenker for interesting conversations related
with this work.

The present work is supported by NSF Grant No. PHY89-04035.
The work of A. J. is partially supported by the Department of Energy
under contract DE-AC02-76ERO3130.A021-Task A.
The work of T. Y. is partially supported by
Grant-in-Aid for Scientific Research on Priority Areas ``Infinite
Analysis'' (No. 04245208) and Grant-in-Aid for Scientific Research
(No. 04640283) from the Ministry of Education, Japan.


\end{document}